\documentclass{aa}
\usepackage{psfig}
\begin{document}


%
   \title{The INTEGRAL\thanks{INTEGRAL is  an ESA project with instruments and science data centre funded by ESA member states (especially
the PI countries: Denmark, France, Germany, Italy, Switzerland, Spain), Czech Republic and Poland, and with the participation of Russia
and the USA. } Burst Alert System}


   \author{S. Mereghetti\inst{1}, D. G\"{o}tz\inst{1}$^{,}$\inst{2},
    J. Borkowski\inst{3},  R. Walter\inst{3}, H. Pedersen\inst{4}}

   \offprints{S. Mereghetti, email: sandro@mi.iasf.cnr.it}

   \institute{Istituto di Astrofisica Spaziale e Fisica Cosmica -- CNR,
              Sezione di Milano ``G.Occhialini'',
          Via Bassini 15, I-20133 Milano, Italy
          \and
             Dipartimento di Fisica, Universit\`{a} degli Studi di Milano Bicocca,
             P.zza della Scienza 3, I-20126 Milano, Italy
         \and
             Integral Science Data Centre, Chemin d'\'{E}cogia 16, CH-1290 Versoix, Switzerland
         \and
           Copenhagen University Observatory, Juliane Maries Vej 30,
              DK 2100 Copenhagen,      Denmark
              }


\abstract{
We describe the  INTEGRAL Burst Alert System (IBAS): the automatic software for
the rapid distribution of the coordinates
of the Gamma-Ray Bursts detected by INTEGRAL.
IBAS is implemented as a ground based system, working on the
near-real time telemetry stream.
During the first six months of operations, six GRB have been detected
in the field of view of the INTEGRAL instruments and localized by IBAS.
Positions with an accuracy of a few
arcminutes are currently  distributed by IBAS to the community for follow-up
observations within a few tens of seconds of the event.
\keywords{Gamma Rays : bursts}
}

\authorrunning{S. Mereghetti et al.}

\maketitle

%

\section{Introduction}

For many years after their serendipitous discovery,
Gamma-Ray Bursts (GRB) were relegated as  a puzzling phenomenon
in the field of high-energy astronomy.
The real  breakthrough in their understanding came with the discovery of
X--ray   (\cite{costa}), optical (\cite{vanpa}), and radio (\cite{frail})
afterglows.
This finally allowed to set a
distance scale, proving that at least long ($>$2 s)
GRB are located at cosmological distances and
associated to the final evolutionary stages  of massive stars (\cite{hjorth}).
These findings led to a renewed interest and to enormous developments
in this field during the last few years
(see, e.g., \cite{vpkw}).

It is clear that the rapid derivation and distribution
of accurate sky positions for GRB is crucial to successfully carry out
such studies. Satellite missions specifically devoted to this,
such as \textit{HETE-2} (\cite{ricker}) and \textit{Swift}  (\cite{swift}) have in fact been developed.
Although INTEGRAL is a general $\gamma$-ray astronomy mission, not particularly
optimized for GRB studies, it was soon realized that the unprecedented
imaging performances of its IBIS instrument (\cite{ubertini})
could offer the possibility of rapid localization
of the events observed by chance in its large field of view.
It was therefore proposed to implement
a ''burst alert system'' in order to allow rapid
multi-wavelength  follow-ups (\cite{pedersen}).

Compared to previous and current GRB localization facilities, IBAS represents
a step forward. Error regions at the arcmin level were obtained by \textit{BeppoSAX}
({\cite{sax}})
with typical delays of one hour or more, related to the frequency
(once per 96 min orbit) of the ground contacts. The Inter Planetary Network
(IPN, \cite{ipn}) can provide error regions of a few tens of square arcmin,
but after several hours (or even days).
Real time localizations from CGRO/BATSE were distributed in the past with
Bacodine (now called GCN, \cite{gcn}), but their typical uncertainty was of a few degrees.
Currently, very nice results are being obtained with \textit{HETE-2} (\cite{ricker}).
The GRB positions derived  on-board at the $\sim$degree level are available
within a few seconds, and later (1-2 hours) refined  down to a few arcmin, with a
ground analysis.

A great advantage of the INTEGRAL mission is the continuous contact with the ground
stations during the observations, made possible by its high orbit (3 days period).
As shown below, IBAS is currently able to provide small error regions
($\sim4'$ radius) within few tens of seconds from the GRB.

\section{IBAS description}

\subsection{Overall architecture}

The INTEGRAL Burst Alert System (IBAS) is the automatic software
devoted to the rapid detection and localization of GRB
(\cite{mere2}).
Contrary to most other $\gamma$-ray astronomy satellites, no
on-board GRB triggering system is present on INTEGRAL.
Since the data are continuously transmitted  without important delays,
the search for GRB is done  at the  INTEGRAL Science Data
Centre (ISDC,  Courvoisier et al. 2003).
The fact that IBAS is running on ground has some advantages:
besides the availability of a larger computing power, a very important factor is
the greater flexibility for what concerns software and hardware upgrades, with respect to
systems operating on board satellites.
To take full advantage of this flexibility, the IBAS software
architecture has been designed in a modular way, which allows to plug-in various
programs for the GRB search, based on different instruments and algorithms.

Fig. 1 gives an overview of the  IBAS software architecture.
The telemetry, received at the ESA Mission Operation Center (MOC)
in Darmstadt, is  continuously  transmitted to the ISDC on a 128 kbs dedicated
line. As soon as the data reach  the ISDC, they are processed by the
Near Real Time Data Receipt Subsystem
which  extracts the relevant telemetry packets
and, after some basics checks,  feeds them into IBAS.
IBAS comprises  several independent \textit{Detector Programs}
runing in parallel. They have the task to
trigger on possible GRB and to perform preliminary checks to filter out,
as much as possible, spurious events.
This architecture allows us to use in parallel different methods for the GRB detection,
as well as to run several instances of the same \textit{Detector Program} with different
parameter settings (e.g. timescales, energy cuts, etc...) in order to increase the
sensitivity for  GRB with different properties.

Currently, programs based on two different algorithms using data from the
ISGRI/IBIS detector are in use,
plus one program to detect GRB seen by the anticoincidence shield of SPI
(in this case no directional information is available).
Other \textit{Detector Programs} based on data from JEM-X and SPI are under development.

The trigger messages produced by the \textit{Detector Programs} are then analyzed by
the \textit{Ibasalertd} program which  combines them in order to extract
the maximum information to decide on the reality of the GRB.
The details of the  logic of trigger confirmation
can be defined in a very flexible way by means of a set of parameters
involving significance of detection,
tolerance for positional and temporal coincidence, etc...
The \textit{Ibasalertd} program also  converts
detector coordinates to sky positions based on the best available attitude
information, verifies that the triggers are not due to known variable sources,
and  eventually distributes via Internet an \textit{IBAS Alert Packet} containing
the position of the GRB.
When several trigger messages, received at slightly different times, refer to the
same GRB, they are filtered by \textit{Ibasalertd}  in order to distribute
new \textit{Alerts Packets}
only if relevant new information is available (e.g. an improved localization).

All the IBAS processes are multi-thread applications written in C or C++.
They run as daemon processes, which means that they do not perform any
terminal input/output and run in background. IBAS processes perform several
subtasks in parallel, each one handled by a separate thread.
With some unavoidable exception, the subtasks are independent and do not
block each other.

As can be seen in Fig.1, IBAS also comprises visualization and analysis tools
which can be used for off-line examination of the data. In order to
minimize  the reaction time, the off-line analysis is based on data products
directly saved by the IBAS programs.

Finally, all the IBAS programs interface with the ISDC Alert Management System,
which is used by the ISDC operators and scientists on duty
to monitor the correct functioning
of the software and to react to possible problems and/or interesting scientific
events.

\begin{figure}
      \psfig{figure=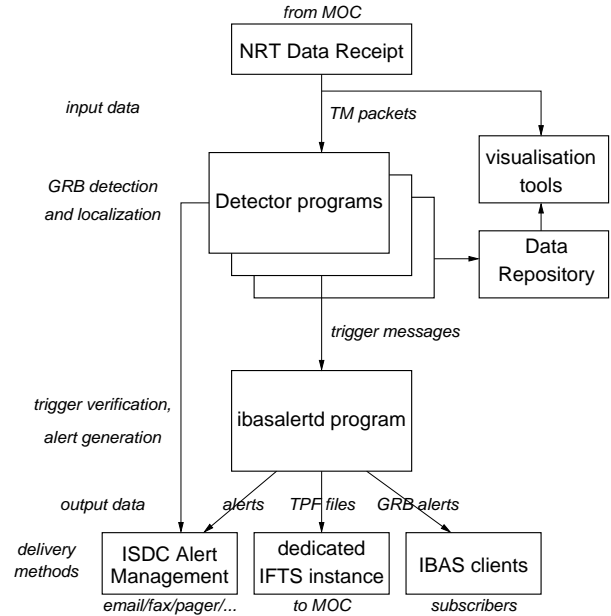,width=8cm,angle=00
      }
      \caption[]{Main components an interfaces of the IBAS software.}
   \end{figure}

\subsection{Detector programs}

Among the INTEGRAL instruments, IBIS
is the most appropriate for GRB localization,
thanks to its large field of view (29$^{\circ}\times$29$^{\circ}$) and
its capability to locate sources at the arcminute level (\cite{gros}).
As mentioned above, IBAS localizations are based on two
different programs using the data from the IBIS lower energy detector ISGRI
(\cite{lebrun}).

Since imaging analysis is the most time consuming part of the algorithm, the
first program performs a simple monitoring of the overall ISGRI counting rate.
This is done by  looking for  significant excesses with respect to
a running average, in a way similar to traditional  triggering algorithms
used on-board previous satellites.
Different instances of this program are currently  running with trigger
timescales ranging from 2 ms to 5.12 s.
The imaging analysis is done only when  a significant counting rate excess is detected.
Images are accumulated for different  time intervals,
deconvolved with very fast algorithms, and compared to the pre-burst
reference images in order to detect the appearance of the GRB as a new source.
This step is essential to eliminate many triggers due to
instrumental effects and background
variations which do not produce a
point source excess in the reconstructed sky images.

The algorithm used in the second \textit{Detector Program} is entirely based on image comparison.
Images of the sky are continuously produced and compared with the previous ones to search
for new sources. With respect to the other program,
this one has the advantage of being less affected by variability of
the background or of other sources in the field of view.
Currently, this program is sampling timescales from 10 to 40 s.

Finally a third kind of \textit{Detector Program} is used to search for GRB detected
by the Anti Coincidence System (ACS) of the SPI instrument. The available data consist of
light curves with the overall ACS count rate binned at 50 ms resolution
(\cite{vonkienlin}).
Although no directional information is available, the resulting triggers can in principle
be used by the \textit{Ibasalertd} program to confirm low significance events seen in other
instruments.

\subsection{Distribution of the IBAS Alert Packets}

IBAS \textit{Alert Packets} with the GRB information are sent via Internet,
using the UDP transport protocol.
Each packet is 400 bytes long, and consists of several fields, the
format and content of which is explained in detail in the documentation available
at the ISDC web pages
\footnote{http:$//$isdc.unige.ch}.
Different types of \textit{Alert Packets} are distributed. Users can select which type(s)
they want receive.
Users interested in receiving the IBAS \textit{Alert Packets}
can also download a \textit{Client Software}, written in standard C language
and tested on the most popular operating systems, that allows them to receive the
\textit{Alert Packets} and to easily use their content, e.g.  in the software commanding
robot telescopes.

IBAS can send more than one packet for each GRB. After the
first one (WAKEUP type) distributed with the shortest delay,
one or more packets of type REFINED are sent automatically if a more
precise localization becomes available.
Finally, one or more packets of type OFFLINE can be sent manually after the
interactive analysis of the data.

Since automated telescopes can exploit the \textit{a priori} knowledge
of the INTEGRAL pointing direction (e.g. to reduce the slew time in case
of a GRB alert, to obtain reference images of the pre-GRB sky, to monitor
the counterparts of INTEGRAL targets), IBAS is also sending packets containing
updated pointing information each time a slew to a new direction begins.
Test \textit{Alert Packets} of each type are sent every 6 hours, to allow the IBAS
users to check their software.

\subsection{Automatic reconfiguration of the Optical Monitor Camera}

INTEGRAL also carries an Optical Monitor Camera (OMC, \cite{mashesse}),
which consists of a 50 mm telescope covering the central
5$^{\circ} \times$5$^{\circ}$ region of the IBIS and SPI field of view.
During normal operations,
owing to the limited  telemetry rate allocated to the OMC,
only the data from a number of small pre-selected windows around sources of interest are
recorded and transmitted to the ground.
The  \textit{Ibasalertd Program}  checks whether the derived GRB position
falls within the OMC
field of view. In such a case, the appropriate telecommand with the definition
of a new window centered on the interesting region is generated and sent to
the MOC to be uploaded to the satellite.
This will allow to quickly observe the GRB/afterglow emission in the optical band.
The OMC observation will consist of several frames with integration times of 60 s
to permit variability studies and to increase the sensitivity for very intense but
short outbursts. The expected limiting magnitude
is of the order of V$\sim$14 for an integration time of 60 s at high Galactic latitudes.

\begin{figure}
      \hspace{0cm}\psfig{figure=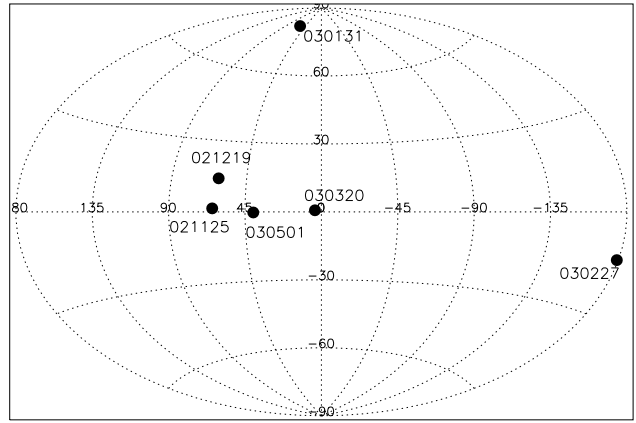,width=8.5cm,angle=00
      }
      \caption[]{Positions in Galactic coordinates of the six GRB
      localized  so far by INTEGRAL. The large fraction of GRB  at low
      Galactic latitudes reflects the non-uniform sky exposure obtained by
      INTEGRAL during the first months of the mission.}
   \end{figure}

\begin{figure}
      \hspace{0cm}\psfig{figure=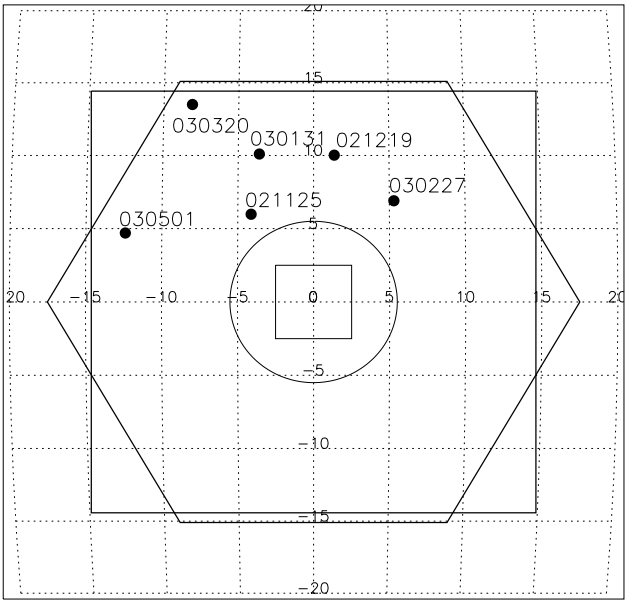,width=8.5cm,angle=00
      }
      \caption[]{Positions of the six GRB in the field of view of the INTEGRAL instruments:
      IBIS (large square), SPI (hexagon), JEM-X (circle) and OMC (small square).
      The scale is in degrees.}
   \end{figure}

\begin{figure}
      \hspace{-1cm}\psfig{figure=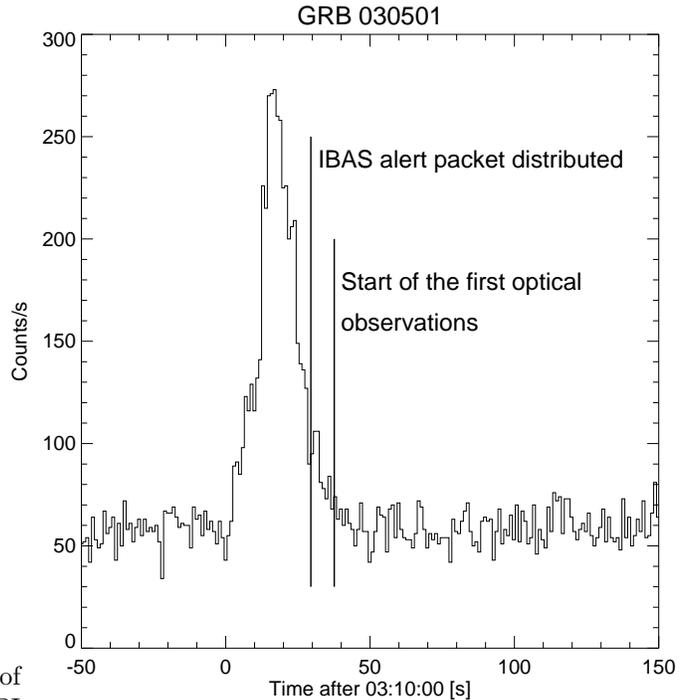,width=9.5cm,angle=00
      }
      \caption[]{Time performance of IBAS in the localization of GRB030501,
      the GRB with the fastest positioning to date.
      The IBIS/ISGRI light curve refers to the 20-200 keV range.
      The two vertical lines indicate the time at which IBAS distributed the
      position with an error radius of 4.4$'$ and the start of the
      first observation of the field obtained with the TAROT
      automatic telescope (\cite{tarot}).}
   \end{figure}

\begin{table*}
\begin{center}
\caption[]{GRB in the IBIS Field of view}
\begin{tabular}{cccccccc}
\hline
\noalign {\smallskip}
GRB & Duration &    Delay$^{a}$ in             & External delivery &  Peak Flux         & Peak Flux     & Fluence & Ref.$^{b}$ \\
    &          & position distribution   & of IBAS &  (20-200 keV)      & (20-200 keV)  & (20-200 keV)   &  \\
    & [s]     &   internal/public    & \textit{Alert Packets} & [ph cm$^{-2}$s$^{-1}$] & [erg cm$^{-2}$s$^{-1}$] & [erg cm$^{-2}$]        &   \\
\hline
\noalign {\smallskip}
021125  &  25  & --$^{c}$ / 0.9 days & OFF & 22  &  2 $\times10^{-6}$  &7.4$\times 10^{-6}$  &  1,2 \\
021219  &   6  & 10 s / 5 hr         & OFF & 3.7 & 3.5$\times10^{-7}$  &  9$\times 10^{-7}$       & 3,4 \\
030131  &  150 & 21 s / 2 hr$^{d}$   & ON  & 1.9 & 1.7$\times10^{-7}$  &  7$\times 10^{-6}$ & 5,6 \\
030227  &  20  & 35 s / 48 min       & OFF & 1.1 & 1.6$\times10^{-7}$  &7.5$\times 10^{-7}$ &   7,8 \\
030320  &  50  & 12 s /  6 hr        & ON  & 5.7 & 5.4$\times10^{-7}$  &1.1$\times 10^{-5}$ &   9,10 \\
030501  &  40  & 30 s / 30 s         & ON  & 2.7 &   3$\times10^{-7}$  &  3$\times 10^{-6}$ &   11,12 \\
 \noalign {\smallskip}
\hline
\label{tab:spec}
\end{tabular}
\end{center}
$^{a}$ Computed from the beginning of the GRB.

$^{b}$ References (for the first announcement and the first journal publication only):
(1) \cite{021125D};
(2) \cite{021125P};
(3) \cite{021219D};
(4) \cite{021219P};
(5) \cite{030131D};
(6) \cite{030131P};
(7) \cite{030227D};
(8) \cite{030227P};
(9) \cite{030320D};
(10) \cite{030320P};
(11) \cite{030501D};
(12) \cite{030501P}.

$^{c}$ The IBAS \textit{Detector Programs} were in idle mode owing to the limited telemetry
allocation for IBIS/ISGRI during this observation. See ref. 2 for details.

$^{d}$ The localization of this  GRB was complicated by the fact that it was
detected while  the satellite was performing a slew between two pointings.
This also reduced its significance level below the threshold for immediate alert distribution.
See Fig. 5 for the resulting error regions.

\end{table*}

\section{IBAS performances}

The IBAS programs have been running almost continuously since
the launch of the INTEGRAL satellite.
The first two months of operations were devoted to the setting of the
many parameters involved in the GRB detection. Some changes in the
algorithms were also required to adapt them to the in-flight data characteristics.
Delivery of the \textit{Alert Packets} to the external clients started on January 17, 2003.
Since then it has always been enabled, except for the period from
February 12 to 28 (during calibration observations of the Crab Nebula),
and for a short interruption (4 hours) on April 23 (for hardware maintenance
reasons).

Six GRB have been discovered to date in the field of view of IBIS
(see Table 1 and figures 2 and 3).
When a GRB is detected by IBAS with high significance, the \textit{Alert Packet}
with the corresponding coordinates is automatically delivered to all subscribed users.
This actually happened so far only for GRB030501. It would have also
occurred for GRB021219 and GRB030227,
but, as mentioned above, they were detected while the external
delivery of alerts was switched off.

Possible GRB detected by IBAS with lower significance generate alerts
reaching only the members of the IBAS Team, who quickly perform an analysis
of the relevant data. If the GRB is confirmed an \textit{Alert Packet}
of type OFFLINE with the derived
position is then distributed. This occurred for GRB030131 and GRB030320.

IBAS has also distributed a number of alerts which were subsequently found
to be unrelated to GRB.
They were retracted by OFFLINE packets and GCN circulars typically
within a few tens of minutes.
These false alerts had different origins. Each of them led to appropriate
changes in the \textit{Ibasalertd} input parameters
and/or to modifications of the software  in order to fix the problem.
Most of the false alerts were caused by an unlucky combination of a
very bright source in the field of view and some unexpected problem
in the instruments and/or ground segment.
In fact, bright (and/or variable) sources are not a problem for IBAS
under normal conditions:
all the triggers are cross-checked with a list of  sky positions
of known bright sources before being distributed.
Obviously, this filter does not work if the wrong source coordinates are computed.
To date this happened a few times, either because the imaging analysis failed due
to some unexpected data configuration related to the instrument or because a
wrong satellite attitude\footnote{
Specifically, problems arise whenever the satellite does not follow the planned timeline
(e.g. due to a telecommand which failed or during  manual  recommanding).
Since the attitude data from the star trackers are transmitted to the ISDC with a delay
of a few  minutes, IBAS must use the  attitude predicted from the timeline
whenever a trigger occurs shortly after the slew (i.e. about every hour, due to the
dithering mode of observation).}
was used to convert from instrumental to celestial
coordinates.

Although in these months we have made considerable progress in the IBAS reliability,
unexpected situations are by their nature difficult to deal with \textit{a priori}
and might generate other false alerts in the future.
IBAS users have the possibility of trading off between speed of reaction time
and reliability of the event by subscribing
only to particular types of \textit{Alert Packets}.

\subsection{Time delay}

The time performances of IBAS for what concerns the GRB detected so far are
summarized in Table 1.

The time delay in the distribution of coordinates results from the sum of
several factors. First of all there is a delay on board the satellite.
This is variable and  depends on the instrument.
In the case of IBIS/ISGRI data the average delay is  about 5 s,
but it can be much longer for other instruments (e.g. approximately 20 s
on average for the SPI ACS data, up to one minute for JEM-X data).
Signal propagation to the ground station is negligible (maximum $\sim$0.6 s),
but some time is required before the data are received at the ISDC passing
through the MOC. This is on average 3 s when the ESA   ground station in Redu (Belgium)
is used, or 6 s when the NASA Goldstone ground station is used.

The time to detect the GRB depends on the algorithm which triggers.
The delay between the trigger time and the GRB onset
is of course dependent on the intensity and  time profile of the event.
The IBAS simultaneous sampling in different timescales should ensure
a minimum delay in most cases, however in practice a minimum of $\sim$5 s is
required to accumulate an image with sufficient statistics.

Finally, the conversion to sky coordinates, comparison with list of
known variable sources, \textit{Alert Packet} construction and delivery require less than
about 2 s.
Of course,  the above numbers assume nominal condition, i.e. no telemetry gaps,
no saturation of the allocated telemetry, no missing
auxiliary data files, etc...

Thus, in many cases, we foresee to be able to generate  alerts
while the  GRB is still ongoing. Indeed this has happened for GRB030501,
whose position
with an uncertainty of only 4.4$'$ reached all the IBAS users   only 30 s after
the beginning of the GRB (Fig. 4).
To our knowledge, such a combination of high speed and small error region
was never achieved before in the localization of a GRB.
Unfortunately, this GRB was located at low Galactic latitude in a region of very
high interstellar absorption which prevented sensitive searches for counterparts
(see Fig. 2).

\begin{figure*}
      \hspace{0cm}\psfig{figure=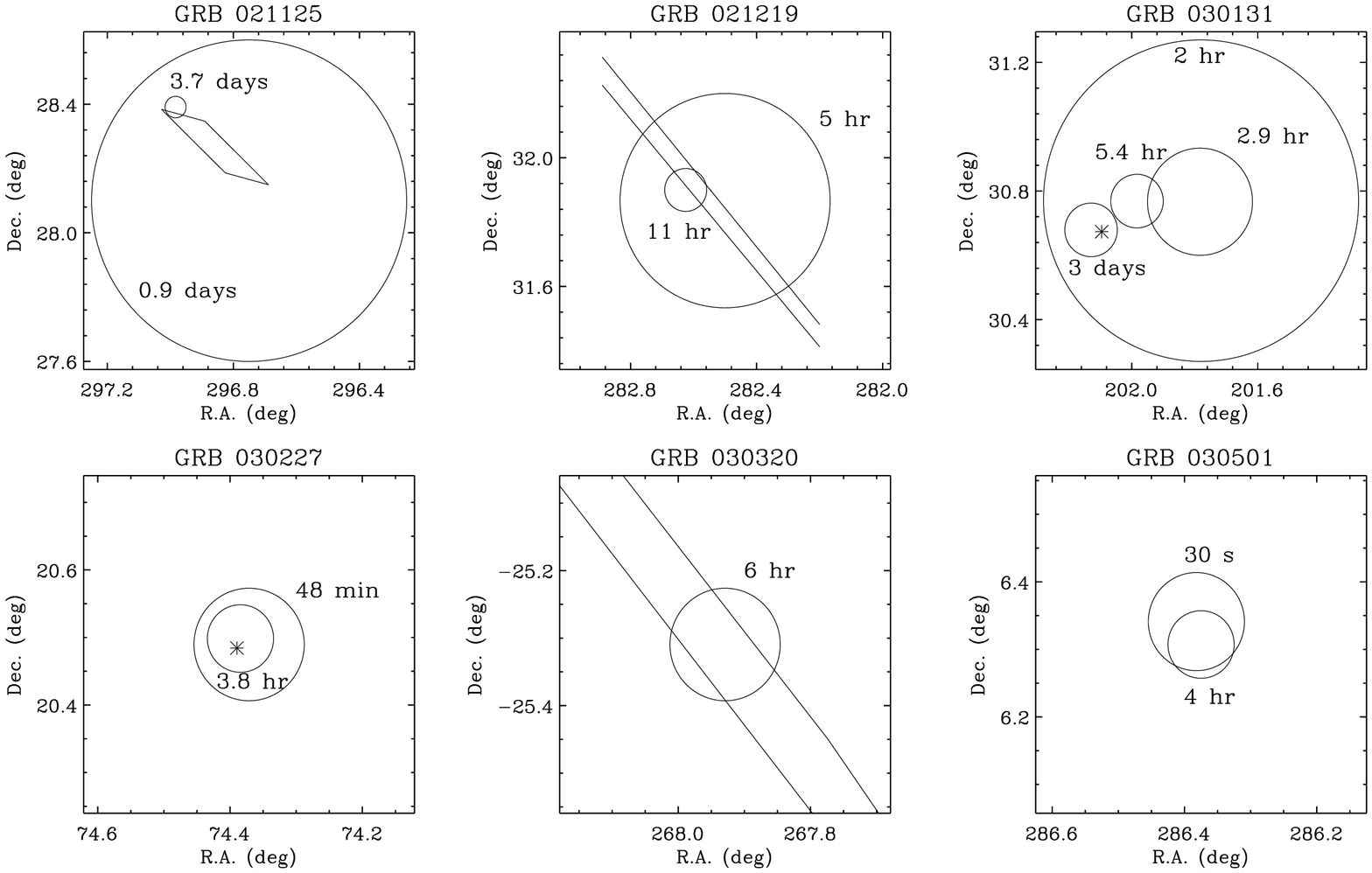,width=17cm,angle=90
      }
      \caption[]{Error regions distributed for the six GRB in the field of view of
      the INTEGRAL instruments, with the corresponding delays.
      Note the different scale of the three upper (1$^{\circ}\times1^{\circ}$) and
      lower (0.5$^{\circ}\times0.5^{\circ}$)      panels.
      The parallelogram and the straight lines indicate error regions independently
      derived with the IPN (\cite{021125IPN,021219IPN,030320IPN}). The asterisks indicate the positions of the
      optical transients associated to GRB030131 (\cite{030131O})
      and GRB030227 (\cite{030227O})
      }
   \end{figure*}

\subsection{Location accuracy}

The source location accuracy (SLA) of coded mask imaging systems
depends on the intrinsic angular resolution of the instrument
and on  the signal to noise ratio of the source. For sources
detected with a high statistical significance the SLA can be a
small fraction of the angular resolution.
The angular resolution of IBIS/ISGRI is $\sim$12$'$, but
sources are typically located with an uncertainty of $\sim$1-2$'$.
For sources detected with a signal to noise ratio of $\sim$30,
the SLA is  smaller than 30$''$ (90\% confidence level).

For most of the time (except, e.g.,  during satellite slews) the
INTEGRAL attitude accuracy is smaller than a few arcseconds  (\cite{walter}).
However, based on the data collected so far, it turns out that the current
uncertainties in the relative alignment between the instruments and
the satellite reference frame, lead to some differences between
the derived and true positions of known sources.
These differences depend on the position in the field of view,
increasing at large off-axis angles. For IBIS/ISGRI these values
range from  15$''$ on-axis to $\sim$2.4$'$ at the border of the field of view.

For this reason, a conservative systematic error of 4$'$ is currently
added in quadrature to the statistical error in the GRB positions
distributed in the automatically delivered \textit{Alert Packets}.
Usually a better accuracy is obtained with the off-line analysis.

Figure 5 summarizes the performance in terms of localization accuracy obtained so far.
Note that at the beginning of the mission the in-flight
instrument misalignment was not calibrated yet.
Therefore,  error radii as large as 20$'$ or 30$'$ were given.
The error regions obtained with the IPN,  and the coordinates
of the optical transients discovered for the two GRB for which prompt observations
could be done, are also shown in the figure. Their agreement with the
INTEGRAL positions confirms that the IBAS localizations are correct.

\subsection{Sensitivity}

The sensitivity of the IBIS/ISGRI detector is very close to that estimated before
the INTEGRAL launch (\cite{ubertini}). Based on such expected sensitivity,
we predicted  a rate of
GRB localizations with IBAS of about one per month (\cite{mcb}), which seems to be
confirmed by the results obtained so far.

A rough evaluation of the sensitivity to GRB can be derived as follows.
The typical IBIS/ISGRI overall count rate in the energy range 15-200 keV used by one of the
IBAS \textit{Detector Programs} varies between about 400 and 800 counts s$^{-1}$,
depending on the background conditions and on the presence of bright sources
in the field of view. For a trigger time interval of 1 s and the current threshold
value, a minimum net count rate of 120-170 counts s$^{-1}$ is required to trigger (and to produce enough
counts to locate the position in the deconvolved image). Assuming a typical GRB
spectrum,
such a count rate corresponds to a flux
of $\sim$0.14-0.22 photons cm$^{-2}$ s$^{-1}$  (20-200 keV).
This applies to the central 9$^{\circ}\times9^{\circ}$
of the IBIS field of view, where the full instrument effective area can
be used. In the external part of the field of view, the so called partially coded
region, the sensitivity is worse.
This explains why the GRB discovered so far have relatively high
peak fluxes compared to the above sensitivity (see Table 1 and Fig. 3).

\section{Conclusions}

The results obtained in the first months of the INTEGRAL mission demonstrate that
IBAS is working as expected. It can provide GRB positions with an accuracy
of a few arcmin within few tens of seconds, at a rate of about one GRB per month.
In addition, IBAS is distributing the light curves of about one GRB per day
detected with the SPI ACS (\cite{vonkienlin}).
These can be used to locate the bursts which
are observed also by other satellites of the IPN.

It is remarkable that,
after only two months from the start of in orbit operation of the instruments,
IBAS was already  functioning successfully,
as demonstrated by the localization of GRB021219.
The results presented here, as summarized in Fig. 5 and Table 1,
indicate that the IBAS capabilities,
in terms of positional accuracy and speed of localization,
have improved during the last few  months.
Although the location in the Galactic plane prevented deep studies of some of the
IBAS GRB, successful observations of optical and X--ray afterglows have been obtained
for GRB030131 (\cite{030131P}) and GRB030227 (\cite{030227P,030227O}).

It is expected that, as more experience is gained with the data and triggering
algorithms, as well as by adding new \textit{Detector Programs} using data
from the other INTEGRAL instruments, the IBAS performances will
improve also in terms of rate of GRB localizations.

\begin{acknowledgements}

The IBAS development has been supported by the Italian Space Agency.
JB acknowledges the Polish State Committee for Scientific Research
for grant number 2P03C00619p02. We thank the members of the INTEGRAL Science
Working Team for their support of the IBAS activities,  several useful
suggestions, and the permission of using PV data for the testing of IBAS.
Davide Cremonesi and Don Jennings gave an important contribution
to the initial development of the IBAS software.

\end{acknowledgements}

\end{document}